# Double Helical Conformation and Extreme Rigidity in a Rodlike Polyelectrolyte

Ying Wang,[1] Yadong He,[2] Zhou Yu,[2] Jianwei Gao,[3] Stephanie T. Brinck,[4] Carla Slebodnick,[1] Gregory B. Fahs,[1] Curt J. Zanelotti,[1] Maruti Hegde,[3] Robert B. Moore,[1] Bernd Ensing,[4] Theo J. Dingemans,[3] Rui Qiao,[2] and Louis A. Madsen*[1]

[1]*Department of Chemistry and Macromolecules Innovation Institute, Virginia Tech, Blacksburg, Virginia 24061, United States*

[2]*Department of Mechanical Engineering, Virginia Tech, Blacksburg, Virginia 24061, United States*

[3]*Department of Aerospace Engineering, Delft University of Technology, Kluyverweg 1, 2629 HS, Delft, The Netherlands*

[4]*Van't Hoff Institute for Molecular Sciences, University of Amsterdam, Science Park 904, 1098 XH, Amsterdam, The Netherlands*

## Abstract

The ubiquitous biomacromolecule DNA has an axial rigidity persistence length of ~ 50 nm, driven by its elegant double helical structure. While double and multiple helix structures appear widely in nature, only rarely are these found in synthetic non-chiral macromolecules. Here we describe a double helical conformation in the densely charged aromatic polyamide poly(2,2'-disulfonyl-4,4'-benzidine terephthalamide) or PBDT. This double helix macromolecule represents one of the most rigid simple molecular structures known, exhibiting an extremely high axial persistence length (~ 1 micrometer). We present X-ray diffraction, NMR spectroscopy, and molecular dynamics (MD) simulations that reveal and confirm the double helical conformation. The discovery of this extreme rigidity in combination with high charge density gives insight into the self-assembly of molecular ionic composites with high mechanical modulus (~ 1 GPa) yet with liquid-like ion motions inside, and provides fodder for formation of new 1D-reinforced composites.

DNA molecules, which act as a storage and transfer platform for the genetic information of life, exhibit a double-stranded helical conformation that has been known for decades.[1,2] Further applications of helical macromolecules involve not only enantioselective and asymmetric catalysts for chemical reactions, but also novel scaffolds and templates for supramolecular self-assembly.[3,4] Inspired by natural multi-stranded helical macromolecules such as polypeptides, collagen and polysaccharides, researchers are engaged in constructing synthetic polymers with a variety of helical conformations. Broadly speaking, the most widely used chemical and physical foundations to build double-stranded macromolecules rely on metal-directed (ligand-containing) self-assembly[5,6] and hydrogen-bonding-driven self-assembly (inter-strand hydrogen-bonding).[7,8] In aromatic polymers, the amide linkage functionality is preferred as it participates in H-bonding. Generally, aromatic cores are *meta-* or *ortho-* linked to enable a twist necessary to form a helix, also known as a helical foldamer.[9-11] Additionally, the presence of polar side groups such as hydroxy (*H-bonding*) or pendant aliphatic moieties (*steric hindrance*) drives helix formation.[12] However, only a few structural motifs enable double helical oligomers – such as peptide nucleic acids,[13] amidinium-carboxylate salt bridges,[7,14,15] and coordination polymers.[5,6,16-18] So far, synthetic polymers with double helical conformations are seldom reported other than isotactic poly(methyl methacrylate) (*it*-PMMA)[19] and some self-assembled oligomers.[7,14,20,21] Double helix formation in aromatic oligoamides has so far been driven only by incorporation of a chiral center or through chiral solvation.[10,12] For example, N-heterocyclic amide oligomers (*meta*-linked) exhibit double helical structures in solution and solid state due to formation of strong intermolecular stacking.[10,22]

Herein, we describe a unique aromatic sulfonated polyamide that possesses the double helical structure, thus leading to an extreme axial rigidity of this polymer chain. The double helix forms for this all-*para*, all-aromatic-backbone macromolecule despite the lack of a chiral center. Moreover, we obtain this polymer using a simple, single-step interfacial polycondensation reaction. We verify formation of this double helix via complementary investigations utilizing X-ray diffraction, $^{23}$Na NMR spectroscopy, and molecular dynamics simulations. This synthetic sulfonated aramid polyanion, poly-2,2′-disulfonyl-4,4′-benzidine terephthalamide (PBDT), can be used to form a unique series of hydrogels[23,24] and ion gels,[25,26] which have displayed great potential as next-generation functional materials for batteries, fuel cells and optical sensors.[23-28] The high rigidity imparted by the double helix drives important properties in these new materials. PBDT is a water-miscible polymer that forms a highly anisotropic lyotropic nematic liquid crystal (LC) phase[28] at concentrations down to exceedingly low values ($C_{PBDT} \geq 0.3$ wt%). We have described a nematic LC model of PBDT aqueous solutions previously, but the detailed molecular configuration of PBDT has not been fully investigated. Herein, we further elaborate a number of similarities between PBDT and DNA molecules in order to highlight the unique characteristics of the double helical conformation of this synthetic rigid-rod polyelectrolyte. Probing the similarities between these two systems provides important insight towards the understanding of double helical systems and feeds into design principles for future discovery of functional materials.[23-25]

**Results and Discussion**

**X-ray diffraction, extreme rigidity, and the PBDT double helix.** X-ray diffraction (XRD) is commonly used to determine the crystalline or semi-crystalline structure of small molecules, proteins and macromolecules. In 1953, the DNA double helical configuration was first proposed by Watson and Crick[1,2] based on an XRD pattern of DNA fibers by Rosalind Franklin.[29] Following a similar approach, we also employ XRD to study the packing structure and morphology of PBDT aqueous solutions. Since it is difficult to obtain aligned PBDT fibers, we have directly run XRD experiments on concentrated and magnetically oriented PBDT aqueous solutions. **Figure 1** summarizes the structural configuration of PBDT and our XRD results, including data and simulations. As shown in **Figure 1a**, we observe a highly ordered X-ray diffraction pattern from a 20 wt% PBDT aqueous solution after placing the sample in a $B$ field ($\geq$ 0.5 T). **Figure1b** shows the corresponding simulated results. The strong monodomain orientational order maintained after removal of the sample from the $B$ field indicates the extremely long persistence length of the PBDT chain,[28] which we attribute to the relatively rigid PBDT backbone and more importantly to the double helical structure.

Indeed, based on the concentration above which the aligned phase forms (1.5 wt%)[28] the persistence length (stiffness) along the PBDT rod axis[30] is > 240 nm based on Onsager theory,[31] and ~ 670 nm based on Flory theory[32] (see also Supplementary Information). This persistence length is substantially longer than DNA (~ 50 nm)[33] and other rigid helical molecules such as poly(benzyl-l-glutamate) (PBLG).[34] Thus PBDT exhibits perhaps the highest known rigidity persistence length of any simple molecular structure. With refinements in the synthesis of PBDT beyond our original study, yielding a higher molecular weight of $M_w \approx$ 180 kg/mol (see SI for details), we have observed that the aligned phase can form at concentrations down to 0.3 wt% PBDT in water, which represents a persistence length of > 1.2 μm.

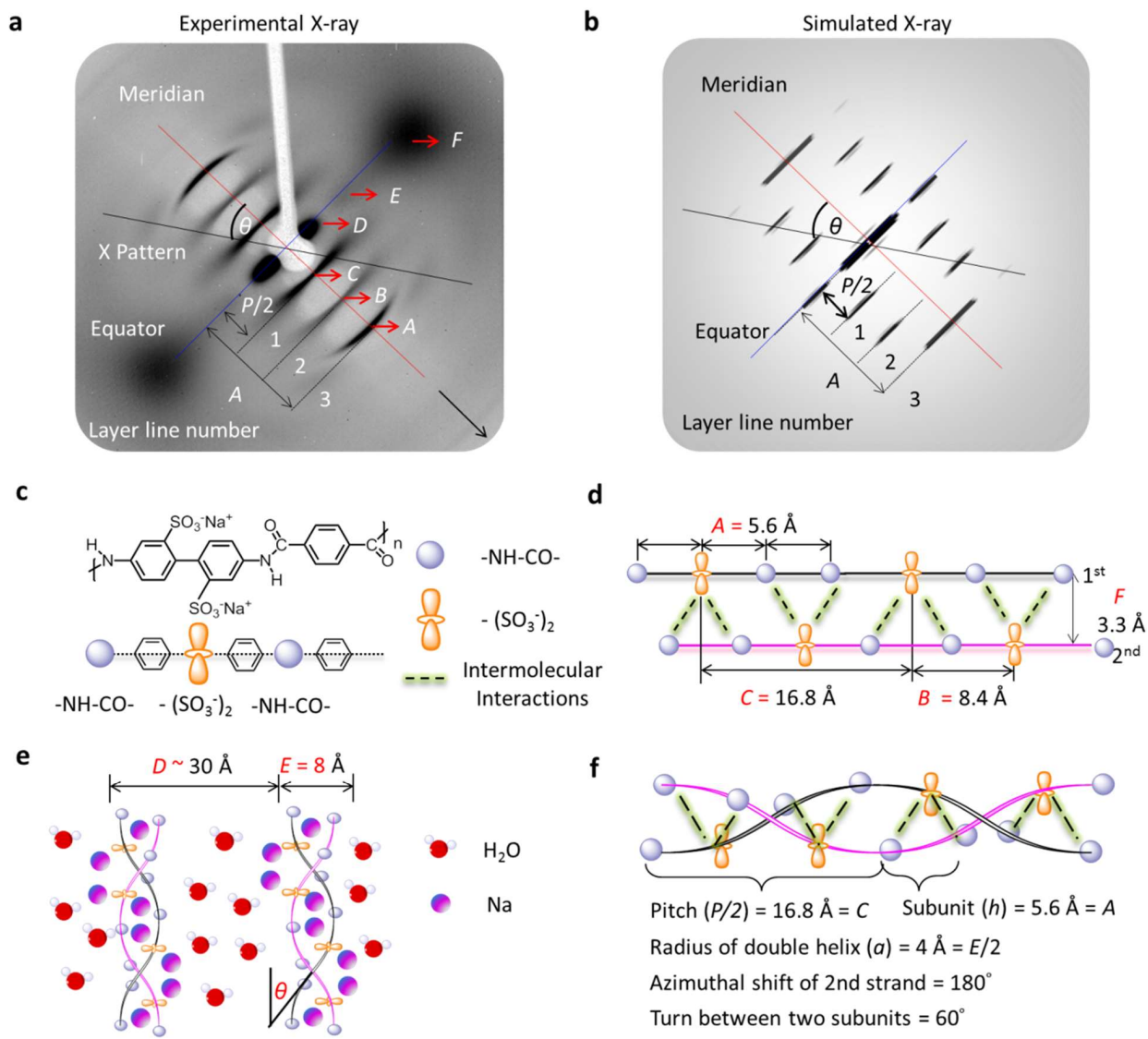

**Figure 1.** Structural motifs along with experimental and simulated XRD results for 20 wt% PBDT aqueous solution. (a) The X-ray diffraction pattern for 20 wt% PBDT aqueous solution. The main diffractions are labelled with *A*, *B*, *C*, *D*, *E* and *F* and the helix tilt angle is *θ*. (b) The simulated X-ray diffraction pattern based on the "HELIX" software package. The layer-lines 1, 2 and 3 in the simulated results clearly mimic the experimental results. (c) The chemical repeat unit of PBDT includes one set of $-SO_3^-$ groups (2 sulfonate groups from one biphenyl unit) and two –NHCO– groups, each of which are mutually connected by one benzene ring. (d) The 2$^{nd}$ PDBT strand is shifted 8.4 Å ($B = P/4$) away from the 1st strand along the helix axis. Numerous intermolecular interactions between chains (notably hydrogen bonding, dipole-dipole, and/or ion-dipole interactions between $-SO_3^-$ and $-NHCO-$ groups – shown as green dashed lines) and the rotation of each subunit contribute to the double helical conformation. (e) PBDT double helices self-assemble into an aligned (nematic) morphology in aqueous solution. The purple dots refer to $Na^+$ counterions. The red dots are water molecules. (f) Elucidation of inter-chain bonding and molecular packing for PBDT with the double helical conformation. The helical parameters used for the X-ray simulation are also listed. ($P$ = 33.6 Å, $h$ = 5.6 Å, $a$ = 2 Å, Azimuthal shift = 180°)

In **Figure 1a**, the black arrow on the bottom right shows the direction of sample alignment in the instrument, which is also the aligned axis of molecular orientation, which can be imposed by either a weak applied magnetic field or simply by the cylindrical confines of an X-ray capillary.[28] The red solid line that passes through the diagonal of the pattern is the meridian. The diffraction peaks on the meridian are called meridional reflections, which usually represent subunit axial translations.[35] The blue solid line perpendicular to the meridian is the equator, with peaks called equatorial reflections. A helical diffraction pattern is usually represented by a series of layers or arcs along the meridian, called layer lines.[35,36] The layer line with closest distance to the equator gives the pitch length for a single helix or half pitch length for a double helix such as that described here. Based on the relationship between real space and reciprocal space, the positions of the layer lines are related to successive orders of the pitch length (P), which in this case is 33.6 Å. Starting from the equator, for a double helix, the layer lines are at spacings P/2, P/4, and P/6.[35,36] All of the layer lines are at each side of the meridian, but usually not on the meridian, except for when the layer lines also correspond to the subunit translation.[35,36] Based on this background, the main diffraction patterns in **Figure 1a** are labeled with *A*, *B*, *C*, *D*, *E* and *F* (*A* = 5.6 Å, *B* = 8.4 Å, *C* = 16.8 Å, *D* ≈ 30 Å, *E* = 8 Å and *F* = 3.3 Å). By correlating the diffraction patterns and the helical parameters, we can elucidate the double helical conformation of PBDT chains in aqueous solutions as follows.

In **Figure 1c**, the chemical repeat unit of PBDT includes a set of two $-SO_3^-$ groups that coincide approximately in a plane perpendicular to the molecular axis and two –NHCO– that are mutually connected by one benzene ring. The theoretical chemical repeat unit length of PBDT along the average molecular alignment axis is *ca.* 17.4 Å[27] assuming all bonds are in the trans conformation. Here we observe a repeat distance along the aligned rod axis of 16.8 Å (*C = P/2*). This repeat unit is composed of three subunits with equal length of 5.6 Å (*A = P/6*). **Figure 1d** shows the combination of two PBDT chains, where the 2[nd] strand is shifted 8.4 Å (*B = P/4*) away from the 1[st] strand. Intermolecular interactions such as hydrogen bonding (between the $-SO_3^-$ and –NHCO– functionalities), π stacking, and dipole-dipole interactions along with the rotation of each subunit contribute to the formation of the double helical conformation. *B* is the distance between adjacent $-SO_3^-$ functionalities along the polymer chain. The average inter-helix distance (within one double helix) is represented by *F*. **Figure 1e** shows schematically the self-assembly behavior of PBDT chains in $H_2O$ (red) solution with $Na^+$ counterions (purple). In 20 wt% PBDT aqueous solution, the rod-rod distance is ~ 30 Å, represented by *D*. This spacing is not clearly defined due to the interference of the diffracted beam with the beam stop, but we have further characterized this distance by small-angle X-ray scattering (SAXS), as introduced below. *E* corresponds to the overall average diameter of the double helix. In addition to the main diffraction peaks along the equator and meridian, we explain the displayed layer-line patterns by analogy with the layer-line diffractions of DNA fibers. The symmetry of a helical structure can usually be defined in terms of a number of parameters, including the subunit axial translation ($h$), the pitch of the helix ($P$), and the radius of the helix ($r$).[35,36] The number of subunits in one pitch equals $P/h$. To describe the origin of the layer-line patterns, we propose a model with critical parameters in **Figure 1f**, based on which we simulate a diffraction pattern (**Figure 1b**) using a helical diffraction simulation software package called "HELIX".[36] We include detailed simulation parameters and information about HELIX in Supplementary Information Figure S1. The simulated results show high coincidence with the experimental results. The main difference is the absence of reflections on the meridian along the layer line number 1 and 2. We attribute this discrepancy to two factors: (1) The HELIX software cannot differentiate the $-SO_3^-$ and $-NHCO-$ functionalities, which are represented as the same subunits in the simulation program. (2) We are observing a liquid crystalline aligned phase with a chain orientational order parameter $S = 0.8$,[28] whereas the HELIX program assumes perfect chain ordering as in dry DNA fibers. In other words, the imperfect alignment of the phase results in smearing of peaks (layer lines) into crescents. The additional diffractions on the meridian with length of *B* (*8.4 Å*) and *C* (*16.8 Å*) (**Figure 1b**) are both attributed to axial subunit translation of the $-SO_3^-$ functionalities between two chains and along one chain, respectively. Beyond that, the layer-line spacings in the simulated results agree perfectly with the experimental results, thus confirming the double helical structure of PBDT chains.

**NMR spectroscopy and small-angle X-ray scattering to probe alignment, rigidity, and Na$^+$ counterion associations.** In addition to XRD, quadrupolar NMR spectroscopy provides a complementary measurement of anisotropic structure and molecular dynamics in PBDT aqueous solutions. Strzelecka and Rill have employed $^{23}$Na-NMR to investigate concentrated sodium-DNA aqueous solutions.[37,38] They observed a $^{23}$Na triplet spectrum, which they attributed to the interaction between the Na quadrupole moment and the Na$^+$ electric field gradient (*efg*) in the

anisotropic ambient environment.[37,38] Herein, the quadrupolar splitting $\Delta v_Q$ of the $^{23}$Na nucleus in a uniaxially aligned system can be expressed as

$$\Delta v_Q = Q_p \rho S_{matrix} P_2(\cos \theta_Q) = Q_p \rho S_{matrix} \frac{(3\cos^2 \theta_Q - 1)}{2} \quad (1)$$

where $Q_p$ is the quadrupolar coupling parameter that represents the $^{23}$Na spectral splitting when the nucleus is static and perfectly aligned, $\theta_Q$ is the averaged angle between the principle axis of the Na$^+$ *efg* tensor $V_{zz}$ and the alignment axis of the PBDT matrix.[39] $S_{matrix}$ is the order parameter of the aligned PBDT chain matrix and $\rho$ is the scaling factor representing the interaction between the quadrupolar probe species $^{23}$Na and the host PBDT matrix.[28,40,41] The average angle $\theta_Q$ will be highly dependent on temperature, local asymmetric environment, and concentration of PBDT. The reported magnitude of the quadrupole splitting decreases with increasing PBDT concentration at low temperatures, whereas the splitting increases with concentration at high temperature. These temperature- and concentration- dependent changes in quadrupole splitting $\Delta v_Q$ are consistent with the magnitude of $P_2(\cos \theta_Q)$. The quadrupole splitting converges to 0, when $\theta_Q$ is at the "magic angle" = 54.7° with $P_2(\cos \theta_Q) = 0$.[37,38]

By developing the experimental design for PBDT solutions based on these two varying factors, (temperature and concentration), we observe that PBDT and DNA solutions display highly consistent NMR spectroscopic (quadrupole splitting) behaviors. In **Figure 2a**, the quadrupole splitting of $^{23}$Na will approximately converge to 0 at a "null concentration" $C_{PBDT} = C_0 \approx 10$ wt% at 25°C. The rod-rod distance at this concentration is 31.6 Å.[28] We attribute this quadrupole splitting dependence to the fact that Na$^+$ ions can exchange, predominantly with –SO$_3^-$ groups, both along the individual charged rods (PBDT double helices) as well as between –SO$_3^-$ groups on different rods. In other words, Na$^+$ can jump between two rigid PBDT rods in an inter-helical interaction, or Na$^+$ can jump along a single rod in an intra-helical interaction. This process is both space and time averaged. Additionally, both intra-helical and inter-helical interactions are highly correlated to the concentration and temperature of PBDT aqueous solutions. As shown in **Figure 2b**, at low PBDT concentration, the rod-rod distance is large, corresponding to dominant intra-helical Na$^+$ interactions (translations) along the polymer chain. Na$^+$ needs to axially translate 33.6 Å (one pitch length) to arrive at the next Na$^+$ with exactly the same coordination location. As concentration increases, the rod-rod distance decreases, and at the concentration 9.1 wt% we observe the intersection point (blue circle) of the experimental data shown in **Figure 2f** with $\Delta v_Q$ = 0 Hz. This specific point corresponds to isotropically averaged Na$^+$ interactions (inter-helical and intra-helical), as shown in **Figure 2c**. With further concentration increase, we observe that $\Delta v_Q$ increases again, corresponding to the dominance of the inter-helical interaction as compared to the intra-helical interaction as shown in **Figure 2d**. Based on this deduction, when $C_{PBDT}$ = 9.1 wt% with $\theta_Q$ = 54.7°, the ratio ($P/r$) of the pitch length (33.6 Å) to the rod-rod distance $r$ is $1/\sqrt{2}$. At this point where Na$^+$ experiences isotropic averaging, we observe $r = r_{isotropic}$ = 48 Å, as displayed in **Figure 2c**.

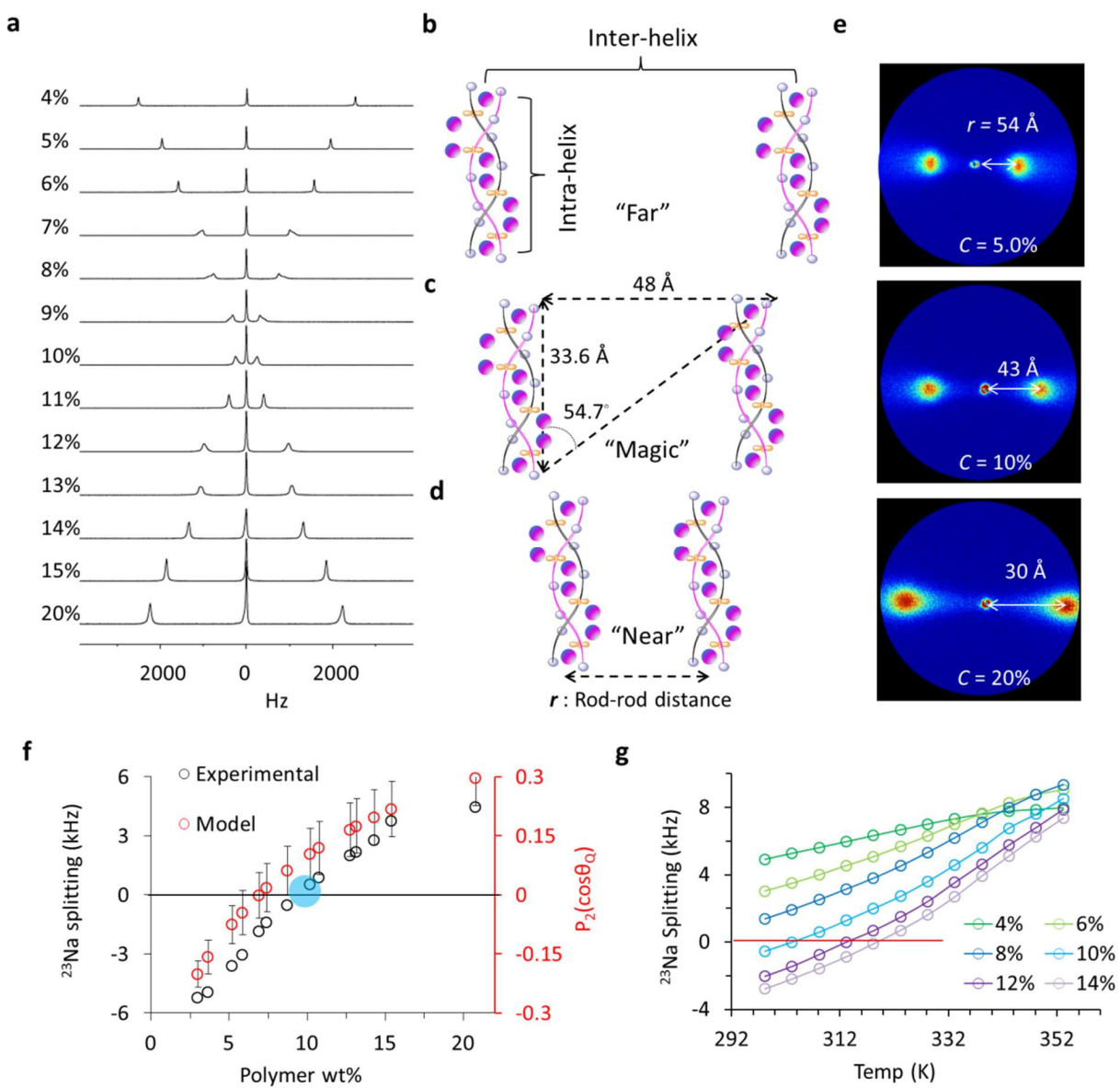

**Figure 2.** $^{23}$Na NMR spectroscopy and SAXS results for PBDT aqueous solutions with varying concentration. (a) shows concentration-dependent $^{23}$Na quadrupole spectra for PBDT solutions at 298 K. (b), (c) and (d) show configurations of PBDT self-assembly behavior with increasing concentration, and the critical geometric parameters at the null point where Na$^+$ shows isotropic dynamics are displayed in (c). Intra-helical Na$^+$ interactions dominate in (b), the null point is at (c) and inter-helical Na$^+$ interactions dominate at (d). (e) shows SAXS results for PBDT aqueous

solutions with $C_{PBDT}$ = 5%, 10% and 20%. In (f), the black open circles represent the observed $^{23}$Na quadrupole splitting as a function of increasing PBDT concentration. The red open circles show the calculated $P_2(cos\theta_Q)$ based on the rod-rod distance (*r*) calculated from the double-stranded hexagonal lattice model and the pitch length (*P = 33.6 Å*) from XRD. (g) shows the $^{23}$Na quadrupole splittings as a function of temperature for PBDT with varying concentrations $C_{PBDT}$ = 4%, 6%, 8%, 10%, 12%, and 14%.

In order to further verify the proposed model, we employ SAXS to investigate the rod-rod distance in PBDT aqueous solutions with varying concentrations. As shown in **Figure 2e**, we can access the rod-rod distance (*r*) based on the scattering vector *q*. By combining the SAXS results and our previously reported double-stranded hexagonal lattice model (Supplementary Information section S3),[28] we use the relationship between rod-rod distance *r* and polymer concentration $C_{PBDT}$ to obtain $r_{isotropic}$ = 43 Å (at $C_{PBDT}$ = $C_0$ = 9.1%), which agrees well with the $r_{isotropic}$ = 48 Å obtained from the proposed model. **Figure 2f** shows the experimental $^{23}$Na splitting and $P_2(cos\theta_Q)$ extracted from the double helix hexagonal lattice model. $\theta_Q$ can be obtained from Equation 2.

$$\theta_Q = \arctan(\frac{r}{33.6}) \qquad (2)$$

The blue area displays the concentration range where the inter- and intra-helical interactions are neutralized (isotropically averaged). Thus, in complement with X-ray scattering, we can utilize $^{23}$Na quadrupolar NMR spectroscopy to quantitatively verify the double helical structure of PBDT.

In addition to the concentration dependence of the $^{23}$Na NMR quadrupole splitting, we can also explore the $^{23}$Na quadrupole splitting as a function of temperature. We observe a similar splitting pattern dependence as with concentration, in which the splitting first decreases across the null point and then increases again with temperature (**Figure 2g**). Furthermore, we observe that $C_0$ is proportional to temperature, as indicated by the red line at $\Delta v_Q = 0$ Hz. This unusual dependence is also clearly observed in aqueous solutions of DNA molecules.[38] The longitudinal ($T_1$) and transverse ($T_2$) spin relaxation times of Na$^+$ counterions in PBDT solutions also parallel results for DNA, and are included in supplementary information section S4.

**Molecular dynamics simulations of double helix formation.** Finally, we describe the PBDT chain configuration based on two MD simulation studies conducted using two separate simulation packages, thermodynamic conditions, and sets of force field parameters. In these MD simulations, we investigate self-assembly of the double helical structure when two or more chains are present. The first set of simulations were done in the NVT ensemble (constant temperature and volume). First, we discuss initial simulation results done in vacuum. We first obtain the optimized structure of the PBDT monomer shown in **Figure 3a**. Starting with polymer chains three monomer units long, we ran simulations lasting 100 picoseconds for two and three chains, with the backbones initially placed at 12 Å distance from each other at $t = 0$ (**Figure 3b**). Multiple configurations for these polymers are initially observed in these gas phase simulations, but substantially before 100 ps, we observe only a self-assembled double helix, as shown in **Figure 3c**. The distance between sulfonate units for the two chains along the self-assembled double helical axis is ~ 8.4 Å (half of the monomer length and ¼ of the helical pitch – see **Figure 1d**), consistent with the above XRD results and model described in **Figure 1**.

In addition to the above model, we also present a second completely distinct simulation result, shown in **Figure 3d**. The estimated pitch length for this model is 3-4 monomer lengths, which is somewhat longer than the first simulation. However, we include this possibility for further investigation, noting that this simulation also shows stable double helix formation even when employing substantially different simulation conditions. In this case, simulations were performed using the Gromacs code in the NVT ensemble.[42] Two PBDT monomers, each with 4 repeating units and measuring ~ 6.8 nm, were initially placed side-by-side in a simulation box filled with water. After energy minimization, the system was equilibrated. After ~ 30 ns, the two PBDT molecules intertwined with each other to form a double helical structure as shown in **Figure 3d.** The system was further equilibrated for ~ 90 ns to ensure the structure is stable. Snapshots of the two PBDT monomers and their self-assembly process are shown in Figures S3 and S4 in Supplementary Information.

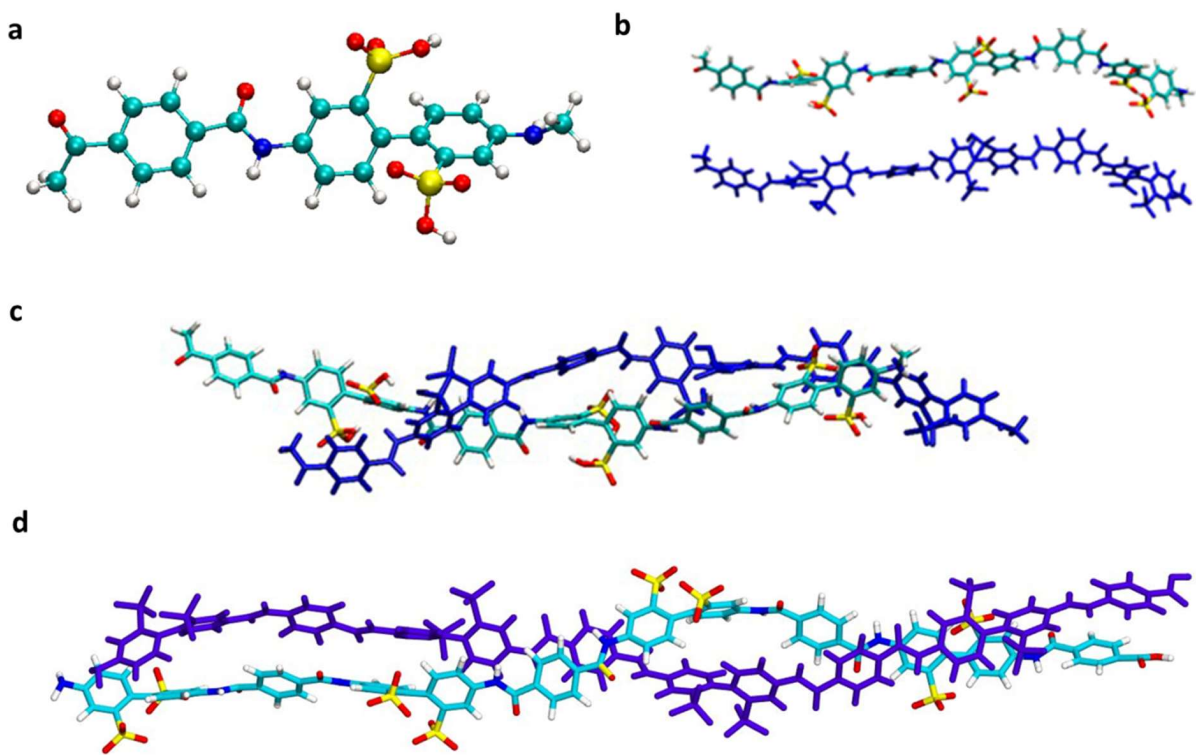

**Figure 3**. Molecular dynamics (MD) simulations of PBDT polymer chains. (a) Optimized monomer structure of PBDT. (b) Simulation result for two oligomers run for 100 ps in vacuum at T = 300 K. (c) Self-assembled double helix with a pitch approximately equal to the XRD results and model of Figure 1. (d) Further MD simulation result in water using a different simulation method, which exhibits a double helix with a somewhat longer helical pitch length.

Above all, we emphasize that both MD models confirm the existence of a double helical structure in this rigid-rod polyelectrolyte. For both of these simulation results, it is clear that establishing this double helix relies on a wide variety of intermolecular interactions, most likely including ion-dipole, pi-pi stacking, hydrophobic effects, hydrogen bonding, and dipole-dipole interactions. This subtle self-assembly of a simple and compact molecular repeat unit involves more complex non-covalent interactions than the simple H-bonding found in DNA and yet with less molecular complexity than most other known double helical structures. Future studies will further unravel the specific interactions at play in order to provide additional insights for informed molecular design for tailored self-assemblies.

**Conclusion**

In this communication, we describe the molecular configuration of the sulfonated aramid PBDT in aqueous solutions. Based on complementary characterizations, we find a number of similarities between PBDT and DNA molecules and demonstrate the unique double-stranded helical structure of PBDT chains in aqueous solution. The strong self-assembly and extreme rigidity present in the PBDT system arises from a varied and synergistic array of interchain interactions, which are more complex and most likely stronger than the simple H-bonding-driven double helical self-assembly mechanism for DNA. From the perspective of macromolecular structural and morphological understanding, this work provides critical ideas and new directions for comprehending molecular packing and ionic association behavior in rigid-rod polyelectrolytes and other self-assembled structures. Furthermore, this work not only resolves the configuration of the PBDT molecular chain, but also lays the foundation for discovering broader applications based on polyelectrolytes with extreme axial rigidity, such as highly conductive yet stiff molecular ionic composites.[25,26] Finally, this highly charged and rigid double helical structure (and/or its derivatives) could potentially become a new one-dimensional (1D) material additive for more general incorporation into functional composites.

**Methods**

*PBDT solutions.* PBDT bulk polymer was dried under vacuum at room temperature for 2 days. Aqueous solutions with PBDT weight percentages (wt%) ranging from 4 to 20% were prepared by loading the prescribed amount of PBDT and $D_2O$ into 5 mm NMR tubes. $D_2O$ was used as received from Cambridge Isotope Laboratories (~99.9%). All of the sample tubes were flame-sealed immediately to prevent evaporation of water and were equilibrated at 80 °C in a water bath for 1 week to ensure complete dissolution and homogenization.

*$^{23}$Na NMR:* The Bruker Avance III WB 400 MHz (9.4 T) NMR was equipped with a 5 mm axial saddle $^{23}$Na rf coil. A simple pulse-acquire sequence with a 90° RF pulse of 5 μs was applied for all measurements from 25 - 80℃.

*X-ray Diffraction (XRD):* The X-ray experiments were carried out on a Rigaku Oxford Diffraction Xcalibur Nova Single-Crystal Diffractometer equipped with an Onyx CCD detector and a Cu microsource operating at 49.5 kV and 80 mA at room temperature. The 20% PBDT aqueous solution was sealed in a quartz capillary with 1.5 mm diameter and 0.01 mm wall thickness and mounted on the edge of a steel pin. The sample to detector distance was 120 mm, providing the data at a scattering angle 2θ from 5° to 34°. The capillary was oriented at a 45° angle relative to the beamstop in order to minimize beamstop interference in the diffraction patterns. The sample was rotated in 2° increments in phi (i.e. along the fiber axis) to generate a total of 200 images each with 1200 s exposure time, and these images were summed to increase signal-to-noise ratio. The software CrysAlisPro (v1.171.37.35, Rigaku Oxford Diffraction, 2015, Rigaku Corporation, Oxford, UK) was used for data collection and analysis.

*Small-Angle X-ray Scattering (SAXS).* All SAXS experiments were operated on a Rigaku S-Max 3000 pinhole SAXS system, equipped with a copper rotating anode emitting X-rays with a wavelength λ of 0.154 nm (Cu Kα) and a sample-to-detector distance of 1600 mm. The relationship between pixel position and scattering vector q was determined by calibrating with a silver behenate standard sample. PBDT aqueous solutions with specific concentrations were sealed in capillaries with 1.5 mm diameter and 0.01 mm wall thickness. Before the SAXS measurements, the capillaries containing polymer solutions were placed axially along the field of a 7.1 T (300 MHz) NMR magnet to achieve uniaxial alignment. Then the capillaries were placed horizontally in the SAXS sample chamber with zero magnetic field. All 2D SAXS patterns were analyzed by the SAXSGUI software package (Rigaku Innovative Technologies, Inc.) to generate the integrated SAXS intensity I($q$) as a function of scattering vector $q$, where $q = (4\pi \sin \theta)/\lambda$. The angle θ refers to one-half of the total scattering angle.

*MD simulations*: In the first MD study, simulations of the polymer chains in vacuum were all accomplished with the CM3D molecular dynamics program. The force field was the Assisted Model Building with Energy Refinement (AMBER), which is often used for simulating proteins and DNA. Periodic boundary conditions (PBC) were used and a Nosé-Hoover thermostat at 300 K was used for simulations at a constant temperature.

The second MD simulation study was performed to investigate the self-assembly of two initially separated PBDT oligomers in aqueous solutions. $Na^+$ counterions were included in the system to ensure electroneutrality. The simulation box was periodic in all three directions and had a size of 6×10×6 nm$^3$. The force field parameters for the PBDT polyanions were generated using the Swissparam server.[43] The TIP3P model was employed for the water molecules. The force fields for the sodium ions were taken from the OPLS-AA force field.[44] The system temperature was maintained at 300 K using the Nosé-Hoover thermostat. The non-electrostatic interactions were computed via direct summation with a cut-off length of 1.2 nm. The electrostatic interactions were computed using the Particle Mesh Ewald (PME) method with a real-space cutoff length of 1.2 nm. All bonds were constrained using the LINCS algorithm.[45]

**Supplementary Information.** This material is available free of charge via the internet.

**Corresponding Author**
*Email:lmadsen@vt.edu

**Acknowledgements**

The authors wish to thank Professors Dimitri Ivanov and Edward T. Samulski for helpful discussions. This work was supported in part by the National Science Foundation under award numbers DMR 0844933, 1507764, and 1507245. Any opinions, findings and conclusions or recommendations expressed in this material are those of the author(s) and do not necessarily reflect the views of the National Science Foundation (NSF).

**Author Contributions**

Y.W., C.S., G.B.F., and R.B.M. designed and conducted X-ray experiments and simulations. Y.W., C.J.Z., and L.A.M. conducted NMR experiments and developed related models.  J.G. and M.H. synthesized the materials.  S.T.B. and B.E. conducted MD simulations in vacuum.  Y.H., Z.Y., and R. Q. conducted MD simulations in water.  Y.W. and L.A.M. developed and compiled concepts and wrote the main paper.  Y.H., Z.Y., J.G., S.T.B., C.S., G.B.F., C.J.Z., M.H., R.B.M, B.E., T.J.D., and R.Q. assisted with compiling and editing the paper.

# Supplementary Information

# Double-Stranded Helical Conformation and Extreme Rigidity in a Rodlike Polyelectrolyte

Ying Wang,[1] Yadong He,[2] Zhou Yu,[2] Jianwei Gao,[3] Stephanie T. Brinck,[4] Carla Slebodnick,[1] Gregory B. Fahs,[1] Curt J. Zanelotti,[1] Maruti Hegde,[3] Robert B. Moore,[1] Bernd Ensing,[4] Theo J. Dingemans,[3] Rui Qiao[2], and Louis A. Madsen*[1]

[1]*Department of Chemistry and Macromolecules and Interfaces Institute, Virginia Tech, Blacksburg, Virginia 24061, United States*

[2]*Department of Mechanical Engineering, Virginia Tech, Blacksburg, Virginia 24061, United States*

[3]*Department of Aerospace Engineering, Delft University of Technology, Kluyverweg 1, 2629 HS, Delft, The Netherlands*

[4]*Van't Hoff Institute for Molecular Sciences, University of Amsterdam, Science Park 904, 1098 XH, Amsterdam, The Netherlands*

**S1: Persistence length (axial rigidity) of PBDT double helix rods**

Based on theories developed by Onsager[1] and Flory[2] that describe the formation of nematic liquid crystalline phases by rodlike particles (which were concisely summarized by Samulski[3]), we can extract the aspect ratio $D/L$ of the rodlike particles and thus the axial rigidity persistence length $L_p$ from the critical volume fraction of rods $\phi_{\text{nematic}}$ at which a nematic phase forms.

Onsager used an athermal model of rods to derive relation 1.[1]

$$\phi_{\text{nematic}} \cong 4.5 \frac{D}{L} \qquad (S1)$$

in which $D$ is the diameter and $L$ is the length of the rods.

In our earlier study of PBDT in solution,[4] on PBDT with $M_w$ = 17,300, we found that the weight fraction at which the nematic phase forms was 0.015 (1.5 wt%). Since the bulk polymer has a density of 1.3 g/cm$^3$, we surmise that the volume fraction of polymer will be even lower than the mass fraction. The aspect ratio $D/L$ would be 300. If we assume that the effective diameter of our double helix rods is 0.8 nm, as given by our XRD studies (the present paper and our previous paper[5]), then the length of the rods will be > 240 nm. This then represents the most conservative value of the the axial rigidity persistence length ($L = L_p$) of PBDT.

Flory later used a lattice model to derive relation-2.[2]

$$\phi_{\text{nematic}} \cong 12.5 \frac{D}{L} \qquad (S2)$$

In this case, the aspect ratio of the rods would be larger (> 830) and the corresponding persistence length is > 670 nm.

With the higher weight average molecular weight $M_w \sim$ 100 kg/mol PBDT that we have synthesized recently (see also below), we have observed that the nematic phase forms at a concentration of 0.3 wt%. This polymer thus yields a persistence length of > 1200 nm based on Onsager theory and > 3300 nm from Flory theory. These findings strongly suggest that the persistence length is at least as long as the contour length of the individual polymer chains, and further suggests that the $L_p$ for these double helical structures may be significantly longer than the values reported here. Indeed, $L_p$ may encompass multiple individual polymer chains that entwine (or interleave) axially to form longer double helices than can be achieved with a simple combination of two chains.

Here we note that molecular weight determination for rigid and charged polymers such as PBDT is rarely reliable (using any available method), and so we are searching for new ways to improve such measurements. As a point of additional information and comparison with our original $M_w$ determination for PBDT stated above,[4] we recently determined the "absolute" molecular weight using a Wyatt MiniDawn LS detector to be 78,000 (with $R_z$ = 34 nm) for our original polymers, and we observed 180k ($R_z$ = 60 nm) for the higher molecular weight polymers that show a N-I transition at 0.3 wt% in water.

**S2: Simulation of X-ray diffraction pattern**

We have employed the "HELIX" simulation package: Version 19-4-04. The "HELIX" package is freely available from (http://www.ccp13.ac.uk ). It is developed by Carlo Knupp and John M. Squire. The detailed information about this software is also available in the software package.

The simulation parameters used in this paper are listed and shown in the HELIX parameter setting interface as follow.

- The axial separation of subunits (h) = 5.6 Å;
- The rotation angle between subunits ($\beta$) = 60°;
- The monomer centre radial position ($a$) = 4 Å;
- The subunit size = 2 Å;
- The azimuthal shift of second strand = 180°;
- The pitch length = h* 360/$\beta$ = 33.6 Å;

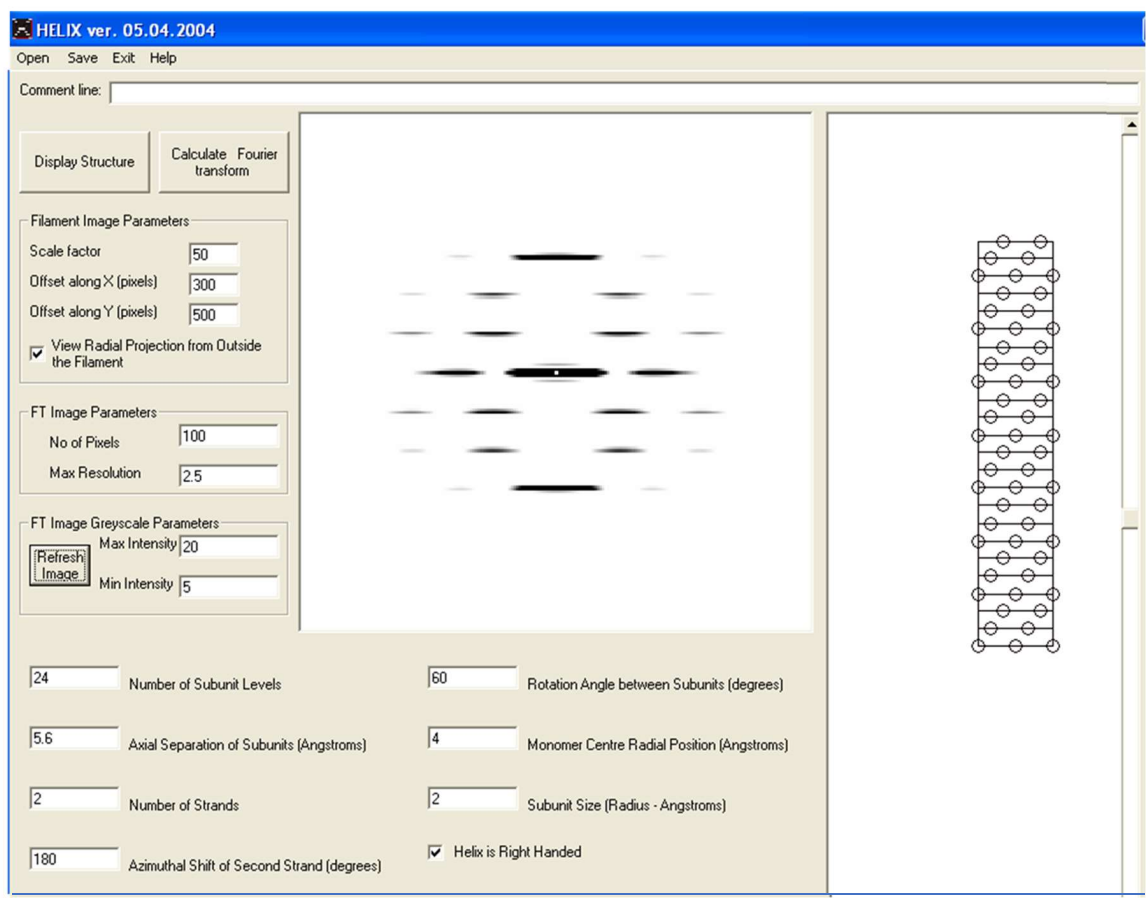

Figure S1. The simulation parameters used in the HELIX to get the simulated X-ray diffraction pattern.

**S3: Rod-rod distance**

According to our previous study[4], we introduced a hexagonal lattice model for PBDT aqueous solutions.[4] We employed a two-parameter least square regression to find the relationship between the rigid rod-rod distance (*r*) and the weight percentage (*C*) of PBDT polymer solutions.

$$r\ (nm) = A * C^{-0.5} + B \qquad (S3)$$

The fitted result is shown as follows with A = 1.0, B = 0.96 and mean square error = 0.078. Compared to the fitting results in the previous study,[4] the equation is updated based on our new SAXS results at higher concentration, including $C$ = 15 wt% and 20 wt%. We have used the procedure defined in reference 4, and have simply used these new fitted A and B values in our analysis of the $^{23}$Na NMR data.

$$r\,(nm) = 1.0 * C^{-0.5} + 0.96 \qquad (S4)$$

### S4: NMR spin relaxation and Na$^+$ dynamics ($T_1$ and $T_2$)

**Figure S2** shows the longitudinal ($T_1$) and transverse ($T_2$) spin relaxation times for Na$^+$ counterions in PBDT polymer solutions. As reported previously by Bull et al., the equality of $T_1$ and $T_2$ for $^{23}$Na in ionic polymer (polyion) solutions originates from fast motions, such as fast internal rotations and librations, counterion exchange, change of direction of the polyion segments, and counterion diffusion along a polyion rod.[6] Similarly, we have not found any apparent difference between $T_1$ and $T_2$ relaxation for $^{23}$Na in PBDT solutions. Thus, we demonstrate that there is fast Na$^+$ motion (fast exchange) along a rod or between rods on at least the $^{23}$Na quadrupolar splitting timescale (< 1 μs). This demonstration lends further support to the proposed model, where there exists a balance, and averaging, between inter-helical and intra-helical Na$^+$ exchanges.[4,5]

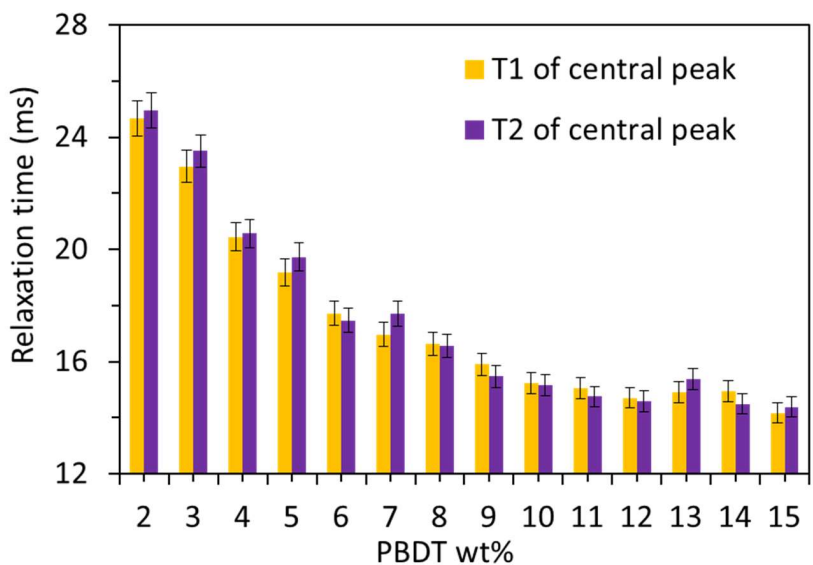

**Figure S2**. $T_1$ and $T_2$ spin relaxation of Na$^+$ counterions in PBDT aqueous solutions as function of the polymer weight percentage in the PBDT polymer solution.

**S5: MD simulation equilibrations and double helix self-assembly**

In Figure S3, after energy minimization, the system was equilibrated. After about ~30 ns, the two PBDT molecules intertwined with each other to form a double helix structure as shown in Figure S4. The system was further equilibrated for ~90 ns to make sure the structure is stable. Figure S5 show the snapshots of the two PBDT monomers at increasing time.

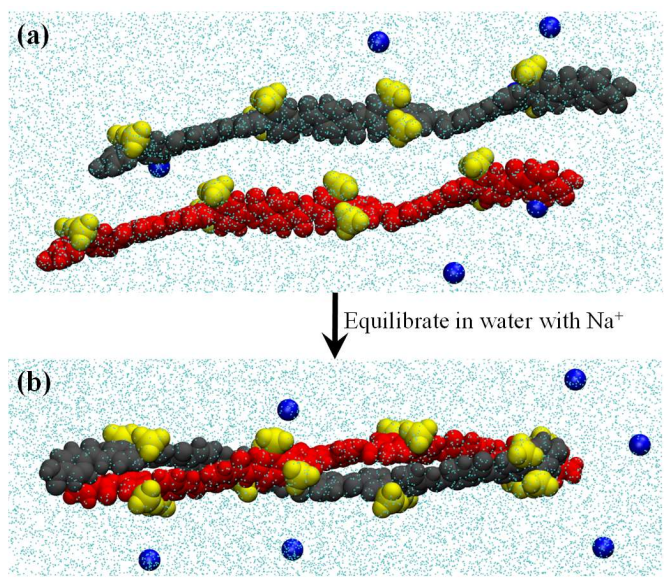

**Figure S3.** Self-assembly of PBDT monomers into a double helix structure. (a) Two PBDT monomers are initially packed side-to-side in water with $Na^+$ ions. (b) After ~30 ns, the two monomers intertwine with each other to form a double helix structure. The red and black balls denote the backbone of the two PBDT monomers, with the yellow balls denote the sulfonate groups. The cyan dots and the blue balls denote the water molecules and the $Na^+$ ions, respectively. Only a portion of the simulation box and the water/ions in the box are shown for clarity.

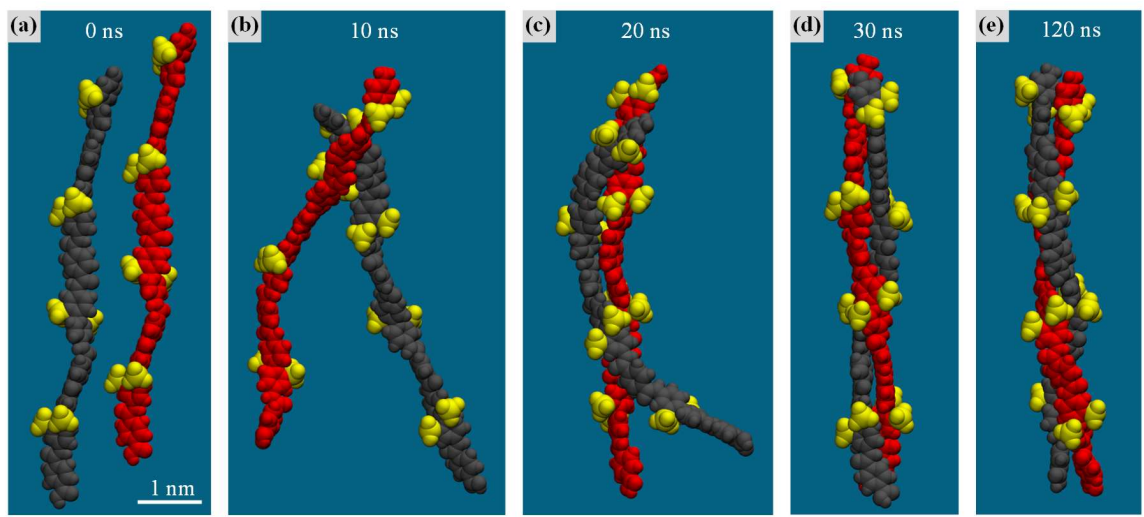

**Figure S4.** Evolution of the conformation of the PBDT monomers during their self-assembly process. The snapshots show the two PBDT polyanions at (a) 0 ns, (b) 10 ns, (c) 20 ns, (d) 30 ns, (e) 120 ns in the simulation. The two PBDT monomers form a double helix structure at ~ 30 ns, and the structure is stable in the rest of the simulation lasting 90 ns. The red and black balls denote the backbone of the two PBDT monomers, with the yellow balls denote the sulfonate groups. The blue background denotes the water molecules. Sodium ions and water are shown for clarity.
8